\documentstyle[preprint,eqsecnum,aps,epsf]{revtex}
\begin{document}
\draft
\title{Level spacing statistics of classically integrable systems\\
-Investigation along the line of the Berry-Robnik approach-}
\author{H.~Makino$^1$ and S.~Tasaki$^2$}
\address{$^{1}$Department of Human and Information Science, Tokai 
University, C-216, 1117 Kitakaname, Hiratsuka-shi, Kanagawa 
259-1292, Japan\\
$^{2}$Department of Applied Physics, Waseda University, 3-4-1 Okubo, 
Shinjuku-ku, Tokyo 169-8555, Japan}
\date{\today}
\maketitle
\begin{abstract}
By extending the approach of Berry and Robnik, the limiting level 
spacing distribution of a system consisting of infinitely many 
independent components is investigated. The limiting level spacing 
distribution is characterized by a single monotonically increasing 
function $\bar{\mu}(S)$ of the level spacing $S$.  Three cases are 
distinguished: (i) Poissonian if $\bar{\mu}(+\infty)=0$, (ii) 
Poissonian for large $S$, but possibly not for small $S$ if 
$0<\bar{\mu}(+\infty)< 1$, and (iii) sub-Poissonian if 
$\bar{\mu}(+\infty)=1$.  This implies that, even when energy-level 
distributions of individual components are statistically 
independent, non-Poissonian level spacing distributions are possible.
\end{abstract}
\section{Introduction}
For a bounded quantum system in the semiclassical limit, statistical 
properties of energy levels at a given energy have been intensively 
studied.  Universal behaviors are found in statistics of 
{\it unfolded} energy levels{\cite{Perc,Berr,Bohi}}, which are 
sequence of numbers uniquely determined by the energy levels using 
the mean level density obtained from the Thomas-Fermi rule(see 
Appendix A).  For quantum systems whose classical counterparts are 
strongly chaotic, the unfolded level statistics are well 
characterized by the random matrix theory which gives level spacing 
distribution obeying the Wigner distribution{\cite{Met,Boh}}.

For quantum systems whose classical counterparts are integrable(such 
systems will be referred to as classically integrable systems), level 
statistics were theoretically studied by Berry-Tabor{\cite{Berr}}, 
Sinai{\cite{Sina}}, Molchanov{\cite{Mol}}, Bleher{\cite{Ble}}, 
Connors and Keating{\cite{CK}}, and have been the subject of many 
numerical investigations.  Still its mechanism is not well 
understood, the level spacing distributions are believed to be the 
Poisson (exponential) distribution for generic classically integrable 
systems.

As suggested e.g., by Hannay (see the discussion of {\cite{Ber}}), 
one possible explanation would be as follows: For an integrable 
system of $f$ degrees-of-freedom, almost every orbit is generically 
confined in each inherent torus, and whole region in the phase space 
is densely covered by invariant tori as suggested by the 
Liouville-Arnold theorem{\cite{4)}}.  In other words, the phase space 
of the integrable system consists of infinitely many tori which have ]
infinitesimal volumes in Liouville measure.  Then, the energy level 
sequence of the whole system is a superposition of sub-sequences 
which are contributed from those regions.  Therefore, if the mean 
level spacing of each independent subset is large, one would expect 
the Poisson distribution as a result of the law of small 
numbers{\cite{Feller}}.

As shown by the pioneering work of Berry and Robnik{\cite{Ber}}, this 
would be the case for the nearly integrable systems consisting of 
large number of independent components.  Quantum energy eigenstates 
are considered to be 'a superposition' of various classical 
trajectories, which are connected by the tunneling effect.  In the 
semiclassical limit, because of the suppression of the tunneling, 
energy eigenstates are expected to be localized on individual region 
in phase space explored by a typical trajectory of classical 
system{\cite{Berr77}}, such as tori and chaotic regions, and to form 
independent components.  The level spacing distribution is, then, 
regarded as a product of superposition of energy levels contributed 
respectively from those independent components.  Based on this view, 
Berry and Robnik observed that the level spacing distribution of the 
system consisting of a single regular component, described by a 
Poisson distribution, and $N$ chaotic components, described by the 
Wigner distributions of equal strength, approaches the Poisson 
distribution when $N\to +\infty$.

The localization of energy eigenfunctions onto individual phase space 
structures was studied by Berry as a semiclassical behavior of the 
Wigner function{\cite{Berr77}}.  For the classically integrable 
system, it is shown that the Wigner function tends to delta function 
on tori in the semi-classical limit{\cite{Berr77-2}}.  Such 
localization phenomenon of eigenfunctions has been observed 
numerically for several systems{\cite{Li,Pro,Mak}}.  For example in 
Ref.{\cite{Mak}}, it was shown that the Husimi distribution function 
of each energy eigenstate is well localized either on the classical 
chaotic regions or on the regular regions filled with classical tori 
and that the energy eigenvalues corresponding to eigenfunctions 
supported by distinct phase-space regions obey distinct statistics. 

Those works seem to imply that the existence of infinitely many 
independent components is an essential ingredient of the appearance 
of Poissonian level spacing distribution and that the semiclassical 
limit is one of the mechanisms providing infinitely many components. 
However, in some classically integrable/nearly-integrable 
systems{\cite{Berr,Ble,Rob,Shni}}, where one might expect infinitely 
many components, deviations from the Poissonian distribution have 
been observed.  Therefore, it is interesting to explore the 
consequences only of the existence of infinitely many independent 
components. 

In this paper, along the line of thoughts of Berry and 
Robnik{\cite{Ber}}, we investigate the distribution of unfolded 
energy levels when energy levels consist of infinitely many 
independent components, and show the possibility of deviation from 
the Poisson distribution.  Hereafter, for the sake of simplicity, 
unfolded energy levels will be referred to as energy levels or levels.

We briefly review the Berry-Robnik theory{\cite{Ber}}.  It relates 
the statistics of the energy level distribution to the phase-space 
geometry by assuming that the sequence of the energy spectrum is 
given by the superposition of statistically independent subspectra, 
which are contributed respectively from eigenfunctions localized onto 
the invariant regions in phase space.  Formation of such independent 
subspectra is a consequence of the condensation of energy 
eigenfunctions on disjoint regions in the classical phase space and 
of the lack of mutual overlap between their eigenfunctions, and, 
thus, can only be expected in the semi-classical limit where the 
Planck constant tends to zero, $\hbar\to 0$.  This mechanism is 
sometimes referred to as {\it{the principle of uniform semi-classical 
condensation of eigenstates}}{\cite{Li94,Robn}}, which is based on 
an implicit state by Berry{\cite{Berr77}}.  The validity of this assumption 
is checked numerically in the semi-classical(high energy) region 
where the Planck volume is much smaller than the phase volume of each 
invariant region{\cite{Li,Pro,Mak2}}.

In the Berry-Robnik approach{\cite{Ber}}, the overall level spacing 
distribution is derived along a line of mathematical framework by 
Mehta{\cite{Met}}, as follows: Consider a system whose classical 
phase space is decomposed into $N$ disjoint regions.  The Liouville 
measures of these regions are denoted by $\rho_i(i=1,2,3,\cdots,N)$ 
which satisfy $\sum_{i=1}^N\rho_i= 1$.  Let $E(S)$ be the gap 
distribution function which stands for the probability that an 
interval $(0,S)$ contains no level.  $E(S)$ is expressed by the 
level spacing distribution $P(S)$ as follows;
\begin{equation}
E(S)=\int_S^\infty d\sigma \int_\sigma^\infty P(x)dx.
\label{eq:1-1}
\end{equation}
When the entire sequence of energy levels is a product of 
statistically independent superposition of $N$ sub-sequences, 
$E(S;N)$ is decomposed into those of 
sub-sequences, $E_i(S;\rho_i)$,
\begin{equation}
E(S;N)=\prod_{i=1}^N E_i(S;\rho_i).
\label{eq:1-2}
\end{equation}
In terms of the normalized level spacing distribution $p_i(S;\rho_i)$ 
of sub-sequence, $E_i(S;\rho_i)$ is given by 
\begin{equation}
E_i(S;\rho_i)= \rho_i\int_S^\infty d\sigma 
\int_\sigma^\infty p_i(x;\rho_i)dx.
\label{eq:1-3}
\end{equation}
and $p_i(S;\rho_i)$ is assumed to satisfy{\cite{Ber}}
\begin{equation}
\int_0^{\infty} S\cdot p_i(S;\rho_i)dS =\frac{1}{\rho_i}.
\label{eq:1-4}
\end{equation}
Note that the spectral components are not always unfolded 
automatically in general even when the total spectrum is unfolded.  
However, in the sufficient small interval 
$[\epsilon,\epsilon+\Delta\epsilon ]$, each spectral component obeys 
a same scaling law (see Appendix {\ref{appendix_a}}) and thus 
is unfolded automatically by an overall unfolding procedure.  
Equations ({\ref{eq:1-2}}) and ({\ref{eq:1-4}}) relate the 
level statistics in the semiclassical 
limit with the phase-space geometry. 

In most general cases, the level spacing distribution might be 
singular.  In such a case, it is convenient to use its cumulative 
distribution function $\mu_i$;
\begin{equation}
\mu_i(S)=\int_0^S p_i(x;\rho_i)dx.
\label{eq:1-5}
\end{equation}
The corresponding quantity of the overall level 
spacing distribution is
\begin{equation}
M(S;N)=\int_0^S P(x;N)dx,
\label{eq:1-6}
\end{equation}

where $P(S;N)$ is the level spacing distribution function 
corresponding to $E(S;N)$.

In addition to equations ({\ref{eq:1-2}}) and ({\ref{eq:1-4}}), we 
assume two conditions for the statistical weights: 
\begin{itemize}
\item
Assumption (i): The statistical weights of independent regions 
uniformly vanishes in the limit of infinitely many regions;
\begin{equation}
\max_i \rho_i \to 0\quad\mbox{as}\quad N\to +\infty.
\label{eq:1-7}
\end{equation}
\item 
Assumption (ii): The weighted mean of the cumulative distribution 
of energy spacing,
\begin{equation}
\mu(\rho;N)=\sum_{i=1}^N \rho_i\mu_i(\rho),
\label{eq:1-8}
\end{equation}
converges in $N\to +\infty$ to $\bar{\mu}(\rho)$
\begin{equation}
\lim_{N\to +\infty}\mu(\rho;N) = \bar{\mu}(\rho).
\label{eq:1-9}
\end{equation}
The limit is uniform on each closed interval: \ $0\le \rho \le S$.
\end{itemize}
In the Berry-Robnik theory, the statistical weights of individual 
components are related to the phase volumes of the corresponding 
invariant regions.  
This relation is satisfactory if the Thomas-Fermi rule for the 
phase space fractions still holds in general.  
Here we do not specify their physical meaning and deal with them 
as parameters.

Under the assumptions (i) and (ii), eqs.({\ref{eq:1-2}}) 
and ({\ref{eq:1-4}}) lead to the overall level spacing distribution 
whose cumulative distribution function is given by the following 
formula in the limit of $N\to +\infty$,
\begin{equation}
M_{\bar{\mu}}(S)=1- \left(1-\bar{\mu}(S)\right) 
\exp{\left[-\int_0^S
\left(1-\bar{\mu}(\sigma)\right)d\sigma \right]}
\label{eq:1-10},
\end{equation}
where the convergence is in the sense of the weak limit.  When the 
level spacing distributions of individual components are sparse 
enough, one may expect $\bar{\mu}=0$ and the level spacing 
distribution of the whole energy sequence reduces to the Poisson 
distribution,

\begin{equation}
M_{\bar{\mu}=0}(S)=1- \exp{\left(-S\right)}.
\label{eq:1-11}
\end{equation}
In general, one may expect $\bar{\mu}\not=0$ which corresponds to a 
certain accumulation of the levels of individual components. 

In the following sections, the above statement is proved and the 
limiting level spacing distributions are classified into three 
classes.  One of them is the Poisson distribution as discussed in 
the original work by Berry and Robnik{\cite{Ber}}.  The others are 
not Poissonian.  We give examples of non-Poissonian limiting level 
spacing distributions in section III. In the concluding section, we 
discuss some relations between our results and other related works.
\section{Limiting Level Spacing Distribution}
\subsection{Derivation of the limiting level spacing distribution}
In this section, starting from eqs.({\ref{eq:1-2}}) and 
({\ref{eq:1-4}}), and the assumptions (i) and (ii) introduced in the 
previous section, we show that, in the limit of infinitely many 
components $N\to +\infty$, the level spacing distribution converges 
weakly to the distribution with the cumulative distribution function:
\begin{equation}
M_{\bar{\mu}}(S)=1- \left(1-\bar{\mu}(S)\right) \exp{
\left[-\int_0^S\left(1-\bar{\mu}(\sigma)\right)d\sigma \right]}
\label{eq:2-1}.
\end{equation}
According to Helly's theorem{\cite{Sin,Feller}}, this is equivalent 
to show that the cumulative distribution function $M(S;N)$ converges 
to $M_{\bar{\mu}}(S)$.  The convergence is shown as follows. 

Following the procedure by Mehta(see appendix A.2 of 
Ref.{\cite{Met}}), we rewrite the gap distribution function $E(S;N)$ 
in terms of the cumulative level spacing distribution functions 
$\mu_i(S)$ of independent components:
\begin{equation}
E(S;N)=\prod_{i=1}^N \left[\rho_i \int_S^{+\infty} 
d\sigma \left(1-\mu_i(\sigma)\right)\right]
=\prod_{i=1}^N \left[1-\rho_i \int_0^{S} 
d\sigma \left(1-\mu_i(\sigma)\right)\right]
\label{eq:2-2}.
\end{equation}
The second equality follows from Eq.({\ref{eq:1-4}}), 
integration by parts and 
$\lim_{\sigma\to +\infty}\sigma \left(1-\mu_i(\sigma)\right)=0$, 
which results from the existence of the average.
Then, the overall cumulative level spacing distribution function 
$M(S;N)$ is given by
\begin{eqnarray}
M(S;N)&=&1+{d\over dS}E(S;N)\nonumber\\
&=& 1- E(S;N) \sum_{i=1}^N {\rho_i-\rho_i \mu_i(S)\over 
1- \rho_i \int_0^{S} d\sigma \left(1-\mu_i(\sigma)\right)}
\label{eq:2-3}.
\end{eqnarray}
First we consider the behavior of $E(S;N)$.  Since the convergence of 
$\sum_{i=1}^N \rho_i \mu_i(\sigma) \to  {\bar \mu}(\sigma)$ for 
$N\to +\infty$ is uniform on each interval $\sigma \in [0,S]$ by 
Assumption (ii),
\begin{eqnarray}
\log E(S;N)&=&\sum_{i=1}^N\log{\left[1-\rho_i\int_0^{S}d\sigma
(1-\mu_i(\sigma))\right]}\nonumber\\
&=&-\sum_{i=1}^N\left[\rho_i\int_0^{S}d\sigma
(1-\mu_i(\sigma))+O(\rho_i^2)\right]\nonumber\\
&=&-\int_0^S d\sigma \left[1-\mu(\sigma;N)\right]
+\sum_i^N O(\rho_i^2)\label{eq:sum}\\
&&\longrightarrow - \int_0^S d\sigma 
\left[1-\bar{\mu}(\sigma)\right]\quad\mbox{as}\quad N\to +\infty
\label{lim:01},
\end{eqnarray}
where we have used $|\mu_i(\sigma)|\le 1$, 
$\log(1+\epsilon)=\epsilon+O(\epsilon^2)$ in $\epsilon\ll 1$, and 
the following property obtained from Assumption (i) 
\begin{equation}
|\sum_{i=1}^N O(\rho_i^2)|\leq C\cdot
\max_i{\rho_i}\cdot\sum_{i=1}^N 
\rho_i =C\cdot\max_i{\rho_i}\to 0
\quad\mbox{as}\quad N\to+\infty.
\label{lim:O}
\end{equation}
with $C$ a positive constant.
The $N\to +\infty$ limit of the sum in the right-hand side of 
equation (\ref{eq:2-3}) can be calculated in a similar way. 
Indeed, as $1/(1-\epsilon)=1+O(\epsilon)$ in $\epsilon\ll 1$, one has
\begin{eqnarray}
\sum_{i=1}^N {\rho_i-\rho_i \mu_i(S)\over 
1-\rho_i\int_0^{S}d\sigma (1-\mu_i(\sigma))}
&=& 1-\sum_{i=1}^N\rho_i\mu_i(S)
+\sum_{i=1}^N O(\rho_i^2)\\
&&\longrightarrow 1-\bar{\mu}(S)
\quad\mbox{as}\quad N\to +\infty.
\end{eqnarray}
Therefore, we have the desired result:
\begin{eqnarray}
\lim_{N\to\infty}M(S;N)&=&M_{\bar{\mu}}(S)
=1- \left(1-\bar{\mu}(S)\right) \exp{\left[-\int_0^S
\left(1-\bar{\mu}(\sigma)\right)d\sigma \right]} 
\end{eqnarray}
\subsection{Properties of the limiting level spacing distribution}
Since $\mu_i(S)$ is monotonically increasing and 
$0\le\mu_i(S) \le 1$, $\bar{\mu}(S)$ has the same properties.  
Then, $1-\bar{\mu}(S)\ge 0$ for any $S\ge 0$ and one has
\begin{equation}
\frac{1}{S}\int_0^S d\sigma 
(1-\bar{\mu}(\sigma))\longrightarrow 
1-\bar{\mu}(+\infty)
\quad\mbox{as}\quad S\to+\infty.
\label{eq:4-7}
\end{equation}
The limit classifies the following three cases:
\begin{itemize}
\item Case 1,~$\bar{\mu}(+\infty)=0$: The limiting level spacing 
distribution is the Poisson distribution.  Note that this condition 
is equivalent to $\bar{\mu}(S)=0$ for ${}^{\forall} S$ because 
$\bar{\mu}(S)$ is monotonically increasing.
\item Case 2,~$0<\bar{\mu}(+\infty)<1$: For large value of $S$, the 
limiting level spacing distribution is well approximated by the 
Poisson distribution, while, for small value of $S$, it may 
deviates from the Poisson distribution.
\item Case 3,~$\bar{\mu}(+\infty)=1$: The limiting level spacing 
distribution deviates from the Poisson distribution for 
${}^{\forall}S$, in such a way that the cumulative distribution 
function approaches 1 as $S\to+\infty$ more slowly than does the 
Poisson distribution.  This case will be referred to as a 
sub-Poisson distribution. 
\end{itemize}
One has Case 1 if the individual level spacing distributions are 
derived from scaled distribution functions $f_i$ as
\begin{equation}
\mu_i(S) =\rho_i\int_0^S 
f_i\left(\rho_i x\right)dx ,
\label{eq:4-8}
\end{equation}
where $f_i$ satisfy 
$$
\int_0^{+\infty} f_i(x) dx = 1,
\int_0^{+\infty} x f_i(x) dx = 1
$$
and are uniformly bounded by a positive constant $D$: $|f_i(S)|\le D$
($1\le i \le N$ and $S\ge 0$).  Indeed, one then has
\begin{eqnarray}
|\mu(S;N)|
&\leq& \sum_{i=1}^N \rho_i^2\int_0^S
\left| f_i\left({\rho_i}x\right)\right|dx \nonumber \\
&\leq& DS\sum_{i=1}^N\rho_i^2
\leq DS\max_i\rho_i\sum_{i=1}^N\rho_i
\longrightarrow 0\equiv {\bar \mu}(S).
\end{eqnarray}
This includes the case studied by Berry and Robnik{\cite{Ber}}, 
where the level spacing distribution is a superposition of a single 
regular component and $N$ equivalent chaotic components, and the 
latter is expressed by the product of the scaled distributions as 
in Eq.({\ref{eq:1-2}}).  Indeed, one has 
\begin{equation}
E^{BR}(S;N)=\exp{\left(-\rho_0 S\right)}
\prod_{i=1}^N E_i^{\mbox{\tiny WIGNER}}(S;\rho_i) \ ,
\label{BRfactor}
\end{equation}
where the statistical weights are $\rho_i = {1-\rho_0\over N}$ and 
the individual level spacing distributions $f_i$ corresponding to 
the gap distributions $E_i^{\mbox{\tiny WIGNER}}(S;\rho_i)$ 
are given by
\begin{equation}
f_i(x)= \frac{\pi x}{2} 
\exp{\left[-\frac{\pi}{4}x^2 
\right]}.
\end{equation}
In addition, this would be the case when the system consists of 
identical $N$ components where the level spacing distribution is 
described by a scaled form as Eq.(\ref{eq:4-8}).  Such a case is 
expected when there is a symmetry such as the regular 
polygonal billiards.  

We remark that, when the limiting function ${\bar \mu}(S)$ is 
differentiable, the asymptotic level spacing distribution admits 
the following density: 
\begin{equation}
P_{\bar{\mu}}(S)= \left[(1-\bar{\mu}(S))^2 
+ \bar{\mu}'(S) \right] \exp{\left[-\int_0^S
\left(1-\bar{\mu}(\sigma)\right)d\sigma \right]}
\label{Density}.
\end{equation}
\section{Example}
\label{sect:exam}
As an example of the deviation from the Poisson distribution, we 
study the quantum systems whose energy levels are described by 
using positive integer numbers, $m$ and $i$ as follows,
\begin{equation}
\epsilon_{m,i} =  m^2 + \alpha\cdot\  i^2,
\label{eq:e}
\end{equation}
where $\alpha$ is the system parameter.  Such energy levels are 
given, for instance, by the rectangular billiard system whose aspect 
ratio of two sides is characterized by $\alpha${\cite{Casati,Rob}}.  
In this paper, the unfolding transformation of the energy levels 
$\{\epsilon_{m,i}\}\to\{\bar{\epsilon}_{m,i}\}$ (see 
Appendix {\ref{appendix_a}}) is done by using the leading 
Weyl term of the cumulative mean number of energy levels,
\begin{equation}
\bar{\epsilon}_{m,i}\equiv\# (\epsilon_{m,i})
=\frac{\pi}{4\sqrt{\alpha}}\epsilon_{m,i}.
\end{equation}
For a given energy interval 
$[\bar{\epsilon},\bar{\epsilon}+\Delta\bar{\epsilon}]$, $i$ or $m$ 
can be regarded as an index which classifies energy levels into 
spectral components.  In this paper, $i=1,2,3,\cdots,N$, 
\begin{equation}
N=\left[\sqrt{\frac{4\sqrt{\alpha}(1+\gamma)
\bar{\epsilon}-\pi}{\alpha\pi}\mbox{ }}\right],
\label{eq:cx}
\end{equation}
is adopted for classification where 
$\gamma\equiv\Delta\bar{\epsilon}/\bar{\epsilon}$, and $[x]$ stands 
for the maximum integer which does not exceed $x$.  The relative 
weight of each component, $\rho_i(i=1,2,3,\cdots,N)$, is given by
\begin{equation}
\rho_i=\left\{\begin{array}{l}
\frac{4(1+\gamma)}{N\pi\gamma (1+\frac{1}{\alpha N^2})}
\left(\sqrt{1+\frac{1-\alpha i^2}{\alpha N^2}}-
\sqrt{\frac{1}{1+\gamma}+\frac{1}{\alpha N^2}
\left( \frac{1}{1+\gamma}-\alpha i^2 \right)}\right)+O\left(
\frac{1}{N^2}\right)\quad\mbox{if}\quad i<\sqrt{\frac{N^2+
\frac{1}{\alpha}}{1+\gamma}},\nonumber\\
\frac{4(1+\gamma)}{N\pi\gamma (1+\frac{1}{\alpha N^2})}
\sqrt{1+\frac{1-\alpha i^2}{\alpha N^2}}+O\left(\frac{1}{N^2}\right)
\quad\mbox{if}\quad \sqrt{\frac{N^2+\frac{1}{\alpha}}{1+\gamma}}
\leq i\leq N.
\end{array}\right.
\end{equation}
As easily seen, $\rho_i$ satisfies the assumption (i);
\begin{equation}
\max_i \rho_i \leq 
\frac{4}{\pi}\sqrt{1+\frac{1}{\gamma}}\times \frac{1}{N}
\left(1+\frac{1}{\alpha N^2}\right)^{-1/2}
+O\left(\frac{1}{N^2}\right)
\longrightarrow 0 \qquad 
\mbox{as}\quad N\to+\infty.
\end{equation}
Note that the limit of infinitely many components, $N\to+\infty$, 
corresponds to the high energy limit, $\bar{\epsilon}\to+\infty$ 
(see Eq.({\ref{eq:cx}})), which is equivalent to the semiclassical 
limit.  In this limit, the statistical weight of each sub-spectrum 
becomes sparse, since each element of $\mu(S;N)$, $\rho_i\mu_i(S)$, 
tends to zero: $\rho_i\mu_i(S)\leq\max_j\rho_j\to0$.

In the billiard system, each spectral component obeys 
a same scaling law (see Appendix {\ref{appendix_a}}), and thus 
is unfolded automatically by an overall unfolding procedure.

Figures 1(a)--1(c) show numerical results of the level spacing distribution $P(S)$ for three values of $\alpha$.  In case that $\alpha$ is far from rational, $P(S)$ is well approximated by the Poisson distribution (Fig.1(a)).  While in case that $\alpha$ is close to a rational, $P(S)$ shows large deviation from the Poisson distribution (Figs.1(b) and 1(c)).  When $\alpha$ is a rational expressed as $\alpha=p/q$, where $p$ and $q$ are coprime positive integers, $P(S)$ is not smooth and becomes a sum of delta functions in the limit of $\epsilon\to+\infty${\cite{Berr,CK,Rob}}, 
which are separated by a same step $X$,
$$
X = \frac{\pi}{4\sqrt{pq}}.
$$

Figures 2(a)--2(c) show the function $|\log{(1-M(S;N))}|$ for three values of $\alpha$ corresponding to figures 1(a)--1(c), respectively.  The dotted line in each figure corresponds to the Poisson distribution, $M(S)=1-exp(-S)$.  As shown in figures 2(b) and 2(c), the distribution function for small value of $S$ clearly deviates from the Poisson distribution.  However, for large value of $S$, it approaches a line whose slope is $1$(see the dashed line), and thus, the level spacings for large value of $S$ obey the Poisson distribution.

In order to compare the non-Poissonian distributions and the classification given in the previous section, we consider,
\begin{equation}
D(S;N)=-\frac{1}{S}\log{\left[1-\int_0^S 
\left(1-M(\sigma;N)\right)d\sigma\right]}.
\end{equation}
When $N\to+\infty$, $D(S;N)$ approaches $\frac{1}{S}\int_0^{S}(1-\bar{\mu}(\sigma))d\sigma$ and this function distinguishes the three cases as follows: In Case 1 i.e., where the level spacing obeys the Poisson distribution, $\lim_{N\to+\infty}D(S;N)=1$.  In Case2, $\lim_{N\to+\infty}D(S;N)\to c$ as $S\to+\infty$ $(c\ne1)$, and in Case 3, where the sub-Poisson distribution is expected, $\lim_{N\to+\infty}D(S;N)\to0$ as $S\to+\infty$.

Figure 3 shows $D(S;N)$ for different values of $N$. From this, one can think that $D(S;N)$ for $N=61905,S\leq 10$ well approximates $\lim_{N\to+\infty}D(S;N)$. 

Figure 4 shows $D(S;N)$ for the three values of $\alpha$ corresponding to figures 1(a)--1(c), respectively.  In case that $P(S)$ is well characterized by the Poisson distribution (figure 1(a)), the corresponding function $D(S;N)$ agrees with $1$.   In case that $P(S)$ deviates from the Poisson distribution ( figures 1(b) and 1(c) ), $D(S;N)$ approaches a number less than $1$ for $S\to+\infty$.  Therefore these results correspond to the Case 2.  In this model, we have not yet observed the clear evidence of Case 3.   Such a case is expected when there is stronger accumulation of the energy levels of individual components.
%
%
\section{Conclusion and Discussion}
In this paper, along the line of thoughts of Berry and Robnik{\cite{Ber}}, we investigated the level spacing distribution of systems with infinitely many independent components and discussed its deviations from the Poisson distribution. In the semiclassical limit, reflecting infinitely fine classical phase space structure, individual energy eigenfunctions are expected to be well localized in the phase space and give independent contribution to the level statistics. Keeping this expectation in mind, we considered a situation where the system consists of infinitely many components and each of them gives an infinitesimal contribution.  And by applying the arguments of Mehta, and Berry and Robnik, the limiting level spacing distribution was obtained which is described by a single monotonically increasing function $\bar{\mu}(S)$ of the level spacing $S$. The limiting distribution is classified into three cases; Case 1: Poissonian if $\bar{\mu}(+\infty)=0$, Case 2: Poissonian for large $S$, but possibly not for small $S$ if $0<\bar{\mu}(+\infty)< 1$, and Case 3: sub-Poissonian if $\bar{\mu}(+\infty)=1$.  Thus, even when energy levels of individual components are statistically independent, non-Poisson level spacing distribution is possible.  

Note that the singular level spacing distribution can be taken into account in terms of non-smooth cumulative distribution $M_{\bar \mu}$.  Such a singularity is expected when there is strong accumulation of the energy levels of individual components.  For certain class of systems, such accumulation is observable.  One example is shown in section III where the results show clear evidence of Case 2.  Another example is studied by Shnirelman\cite{Shni}, Chirikov and Shepelyansky\cite{Shni2}, and Frahm and Shepelyansky\cite{Frahm} for a certain type of systems which contain quasi-degeneracy result from inherent symmetry(time reversibility).  As is well known, the existence of quasi-degeneracy leads to the sharp Shnirelman peak at small level spacings.

One of the interesting features of the level statistics is the level clustering which is described by a nonvanishing value of the level spacing distribution function at $S=0$.  Level clustering is expected for integrable systems or mixed systems, but not for strongly chaotic systems due to the level repulsion.  For certain class of systems, rigorous results are available; Molchanov{\cite{Mol}} analyzed the energy levels of a one-dimensional Schr\"{o}dinger operator with random potential and showed that the level clustering arose from the localization of the eigenfunction in the semiclassical limit.  Minami also analyzed a one-dimensional Schr\"{o}dinger operator with $\delta$-potentials and reported the same result{\cite{Min}}.  The limiting level spacing distribution obtained in this paper possesses level clustering property. Indeed, when $\bar{\mu}(S)$ is differentiable, the level spacing distribution function (\ref{Density}) has nonvanishing value at $S=0$: $P_{\bar{\mu}}(0)= 1 + \bar{\mu}'(0) >0$.

It is also interesting to extend the Berry-Robnik distribution (\ref{BRfactor}) for the level statistics of the nearly integrable system with two degree-of-freedom. The classical phase space of this system consists of regular and chaotic regions.  Since the regular regions corresponding to the system consist of infinitely many independent regions, original proposal for the gap distribution by Berry and Robnik would be replaced by
\begin{equation}
E_{\bar{\mu}}(S;N)=
\exp{\left[-\rho_0 \int_0^S
(1-\bar{\mu}(\sigma))d\sigma\right]}
\prod_{i=1}^{N} 
E_i^{\mbox{\tiny RMT}}(S;\rho_i), \label{eq:4.6}
\end{equation}
where $E_i^{\mbox{\tiny RMT}}(S;\rho_i)$ is the gap distribution function of the $i$th chaotic component obtained from the random matrix theory.  The above distribution formula is classified into cases; Case 1', $\bar{\mu}(+\infty)=0$: Berry-Robnik distribution, Case 2', $0<\bar{\mu}(+\infty)< 1$: Berry-Robnik distribution for large $S$, but possibly not for small $S$, and Case 3', $\bar{\mu}(+\infty)=1$:  A distribution function obtained by the superposition of spectral components obeying the sub-Poisson statistics and the Random matrix theory.  From this classification, one can see that the new formula ({\ref{eq:4.6}}) admits deviations from the Berry-Robnik distribution when $\bar{\mu}(+\infty)\ne 0$.  

For nearly integrable systems with two degree-of-freedom, Prosen and Robnik showed numerically that the Berry-Robnik formula (\ref{BRfactor}) well approximates the level spacing distributions in the high energy region that is called {\it the Berry-Robnik regime}{\cite{regime}}, while it deviates in the low energy region and approximates the Brody distribution.  They studied this behavior in terms of a fractional power dependence of the spacing distribution near the origin at $S=0$, which could be attributed to the localization properties of eigenstates on chaotic components{\cite{fract1,fract2}}.  From the above classification, one can see that the condition $\bar{\mu}(+\infty)=0$ should be satisfied in the Berry-Robnik regime.  While Case 2' and Case 3' in the above classification might propose another possibilities.  When the spectral components corresponding to regular regions show strong accumulation, the level spacing statistics obeys the distribution formula ({\ref{eq:4.6}}) with $0<\bar{\mu}(+\infty)\leq 1$,  and shows deviations from the Berry-Robnik distribution.  Such possibilities will be studied elsewhere.
\vspace{0.2cm}\\
{\large Acknowledgments}\\
\indent
The authors would like to thank Professor M. Robnik, Professor Y.Aizawa, Professor A.Shudo, and Dr G. Veble for their helpful advice.  The authors also thank the Yukawa Institute for Theoretical Physics at Kyoto University. Discussions during the YITP workshop YITP-W-02-13 on "Quantum chaos: Present status of theory and experiment" were useful to complete this work.   This work is partly supported by a Grant-in-Aid for Scientific Research (C) from the Japan Society for the Promotion of Science. 
\appendix
\section{Unfolding of spectrum}
\label{appendix_a}
The unfolding transformation of each energy level  
$\{\epsilon_n\}\to\{\bar{\epsilon}_n\}$ is done by 
using the cumulative mean number of levels up to the 
energy $\epsilon${\cite{Rob}},
\begin{equation}
\#([0,\epsilon])=\int_{0}^{\epsilon}d(x)dx.
\end{equation}
In the above equation, $d(\epsilon)$ exhibits the density of energy 
levels obtained by the Thomas-Fermi rule{\cite{Boh}}:
\begin{equation}
d(\epsilon) = \frac{V(\epsilon)}{(2\pi\hbar)^f},\qquad V(\epsilon)=
\int\delta\left(\epsilon-H({\bf{q}},{\bf{p}})\right)d^{f}{\bf{q}}d^f{\bf{p}},
\end{equation}
where $(2\pi\hbar)^f$ is the Planck volume of the system with $f$ degrees-of-freedom, $V(\epsilon)$ is the phase volume on the energy surface, $\delta(\epsilon -H)$ is the delta function, $H({\bf{q}},{\bf{p}})$ is the classical Hamiltonian function, and $({\bf{q}},{\bf{p}})$ are the coordinates and momenta in the phase space.  The unfolding transformation of spectrum 
$\{\epsilon_n\}\to\{\bar{\epsilon}_n\} (n=1,2,3\cdots)$ is defined 
by
\begin{equation}
\bar{\epsilon}_{n} = \# ([0,\epsilon_{n}]).
\end{equation}

Here, we consider the unfolding procedure of the $i$-th sub-spectrum.  Since the phase volume of the $i$-th component is $\rho_i(\epsilon) V(\epsilon)$, the density of each sub-spectrum is then 
described by
\begin{equation}
d_i(\epsilon)=\frac{\rho_i(\epsilon) V(\epsilon)}{(2\pi\hbar)^f}
=\rho_i(\epsilon) d(\epsilon)
\end{equation}
When the energy interval $[\epsilon,\epsilon+\Delta\epsilon]$ is 
sufficient small, the phase space geometry on the energy 
surface does not change in general.  In other words, $\rho_i(\epsilon)$ 
is approximated by a constant value in this ivterval, and the cumulative 
mean number of the $i$-th sub-spectrum is thus described as
\begin{eqnarray}
\#_i([\epsilon,\epsilon+\Delta\epsilon])&=&\frac{1}{(2\pi\hbar)^f}\int_{\epsilon}^{\epsilon+\Delta\epsilon}\rho_i(e)V(e)de\\
&\simeq&\frac{\rho_i}{(2\pi\hbar)^f}\int_{\epsilon}^{\epsilon+\Delta\epsilon}V(e)de=\rho_i\#([\epsilon,\epsilon+\Delta\epsilon]).
\label{eq_a3}
\end{eqnarray}
Therefore, in the asymptotic limit $\Delta\epsilon\to0$, one can 
see that each spectral component obeys a same scaling law and is 
unfolded automatically by an overall unfolding transformation:
\begin{equation}
\bar{\epsilon}_n =\#(\epsilon_n)
=\frac{1}{\rho_i}\#_i(\epsilon_n).
\end{equation}

The billiard system is very convenient since the phase space 
geometry does not change for variety of $\epsilon$, 
and $\#_i(\epsilon)=\rho_i\#(\epsilon)$ is satisfactory even when the size of 
energy interval does not small.
\newpage
\begin{figure}
\epsfxsize = 15cm
\centerline{\epsfbox{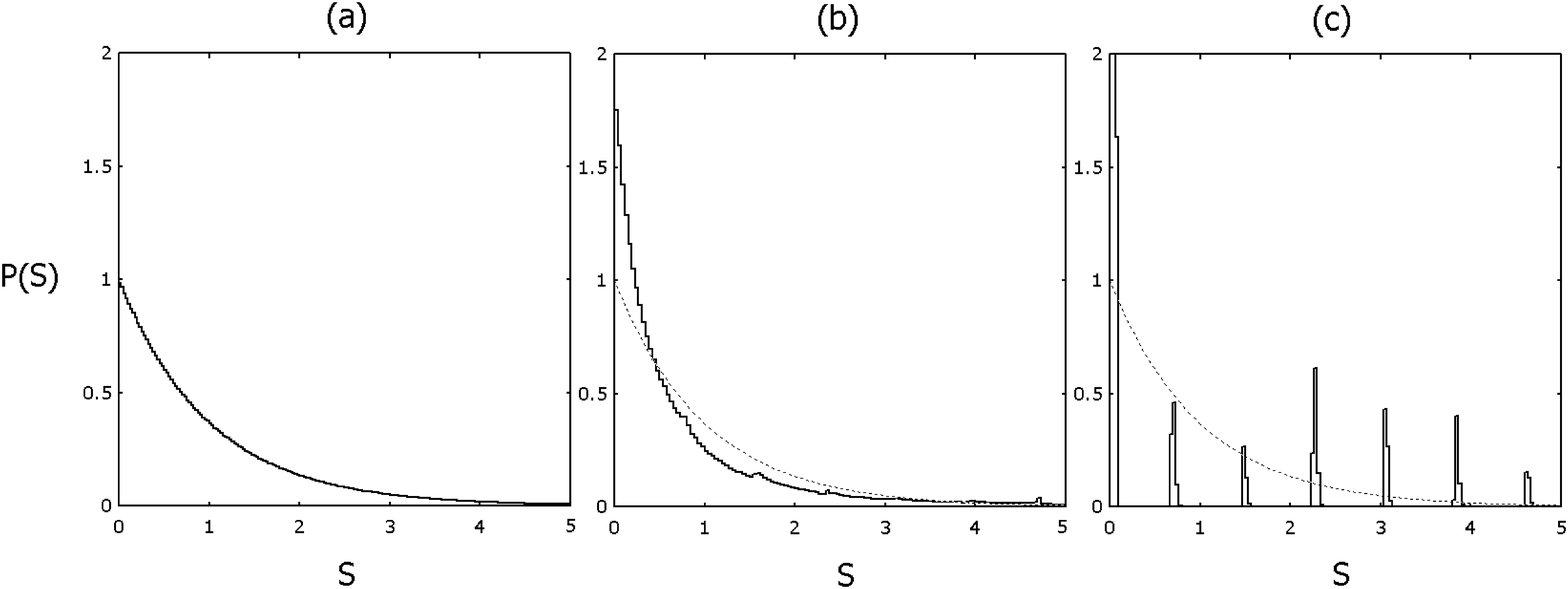}}
{FIG.1  Numerical results of the level spacing distribution 
$P(S)$ for various values of $\alpha$ ;
(a) $\alpha=1+\frac{\pi}{3}\times10^{-4}$,
(b) $\alpha=1+\frac{\pi}{2}\times10^{-9}$, and 
(c) $\alpha=1+\pi\times10^{-11}$.  
We used energy levels $\bar{\epsilon}_{m,i}
\in [300\times10^7,301\times10^7]$.  
Total numbers of levels are (a) 10000016, 
(b) 10000046, and (c)10000043.  The dotted 
curve in each figure shows 
the Poisson distribution, $P(S)=e^{-S}$.}
\label{fig_1}
\end{figure}
\begin{figure}
\epsfxsize = 15cm
\centerline{\epsfbox{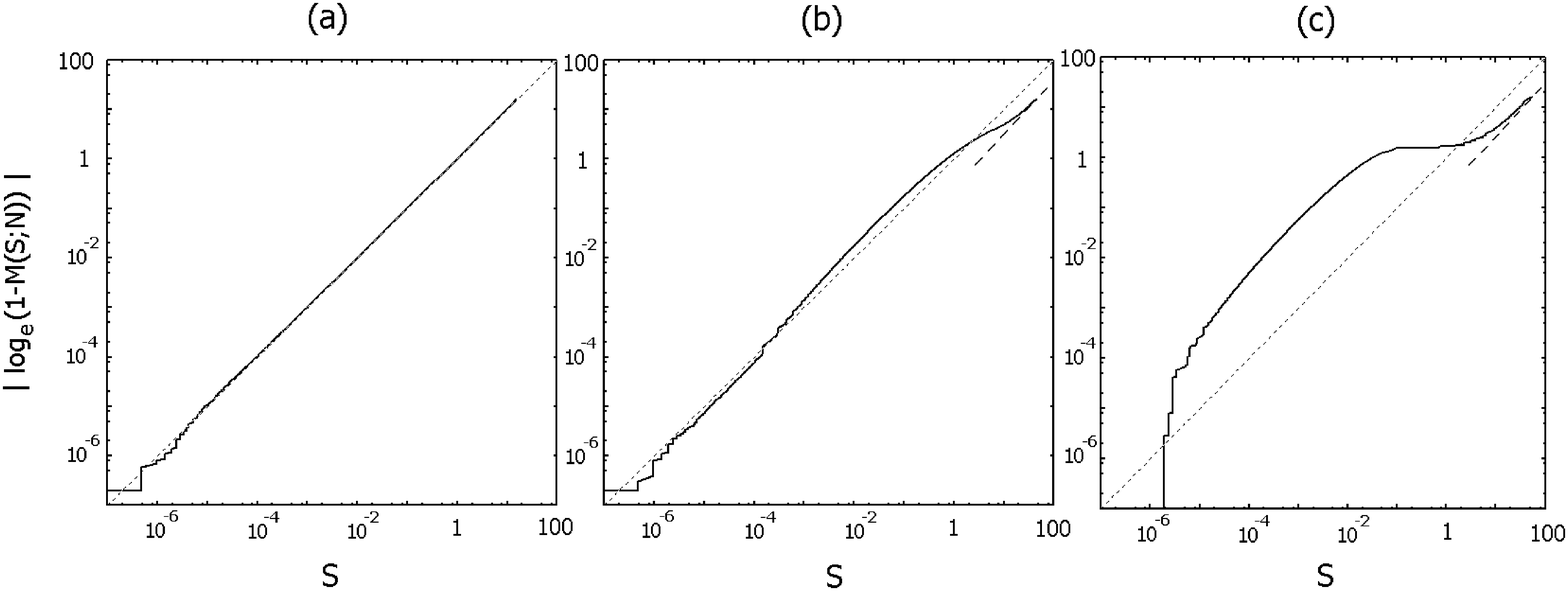}}
{FIG.2  Function $|\log{\left( 1-M(S;N) \right)}|$ for 
(a) $\alpha=1+\frac{\pi}{3}\times10^{-4}$,
(b) $\alpha=1+\frac{\pi}{2}\times10^{-9}$, and 
(c) $\alpha=1+\pi\times10^{-11}$.  
The dotted line in each figure corresponds to 
the Poisson distribution, $M(S)=1-\exp{(-S)}$.}
\label{fig_2}
\end{figure}
\newpage
\begin{figure}
\epsfxsize = 7cm
\centerline{\epsfbox{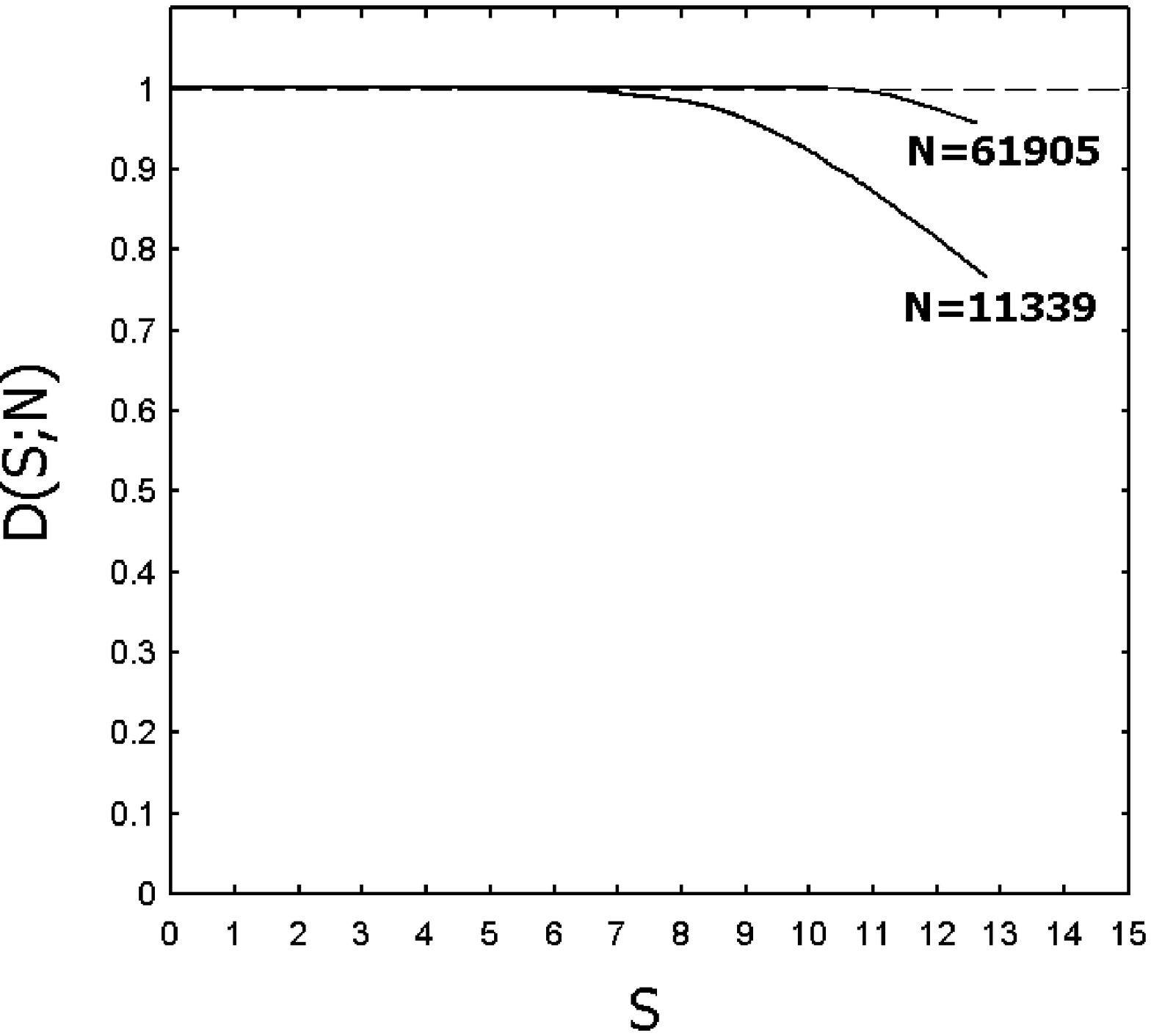}}
{FIG.3  $D(S;N)$ for 
$N=11339$ and for $N=61905$, which correspond to energy levels $\bar{\epsilon}_{m,i}
\in [100\times10^6,101\times10^6]$ 
and $\bar{\epsilon}_{m,i}\in [300\times10^7,301\times10^7]$, respectively.  
In each case, we fixed 
$\alpha=1+\frac{\pi}{3}\times 10^{-4}$.  
The dashed line, $D(S)=1$, exhibits the Poisson distribution.}
\label{fig_4}
\end{figure}
\begin{figure}
\epsfxsize = 7cm
\centerline{\epsfbox{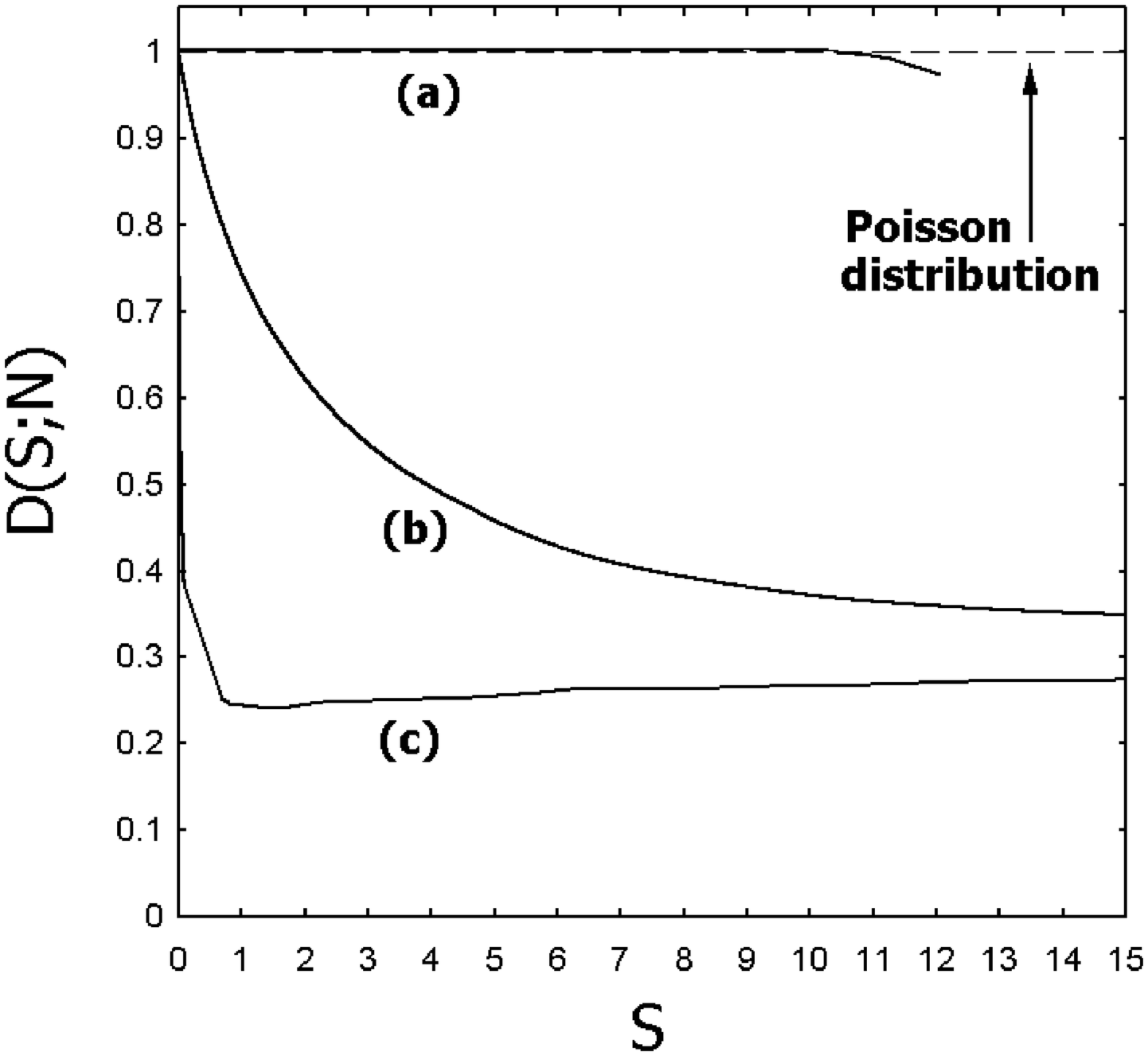}}
{FIG.4  $D(S;N)$ for 
(a) $\alpha=1+\frac{\pi}{3}\times10^{-4}$ and $N=61905$, 
(b) $\alpha=1+\frac{\pi}{2}\times10^{-9}$ and $N=61906$, and 
(c) $\alpha=1+\pi\times10^{-11}$ and $N=61906$, which are calculated from energy levels $\bar{\epsilon}_{m,i}\in [300\times10^7,301\times10^7]$.  The dashed 
line, D(S)=1, exhibits the Poisson distribution.}
\label{fig_3}
\end{figure}
\end{document}